\def\spose#1{\hbox to 0pt{#1\hss}}
\def\lta{\mathrel{\spose{\lower 3pt\hbox{$\mathchar"218$}}
     \raise 2.0pt\hbox{$\mathchar"13C$}}}
\def\gta{\mathrel{\spose{\lower 3pt\hbox{$\mathchar"218$}}
     \raise 2.0pt\hbox{$\mathchar"13E$}}}
\def\ge{\mathrel{\spose{\lower 3pt\hbox{$-$}}
     \raise 2.0pt\hbox{$\mathchar"13E$}}}
\def\le{\mathrel{\spose{\lower 3pt\hbox{$-$}}
     \raise 2.0pt\hbox{$\mathchar"13C$}}}

\documentclass[aps,twocolumn]{revtex4}

\usepackage{graphics}

\begin{document}

\bibliographystyle{apsrev}

\title{LISA data analysis I: Doppler demodulation}
\author{Neil J. Cornish}
\affiliation{Department of Physics, Montana State University, Bozeman, MT 59717}
\author{Shane L. Larson}
\affiliation{Space Radiation Laboratory, California Institute of Technology, Pasadena, CA 91125}

\begin{abstract}
The orbital motion of the Laser Interferometer Space Antenna (LISA) produces amplitude, phase
and frequency modulation of a gravitational wave signal. The modulations have the effect of
spreading a monochromatic gravitational wave signal across a range
of frequencies. The modulations encode useful information about the source location and
orientation, but they also have the deleterious affect of spreading a signal across a wide
bandwidth, thereby reducing the strength of the signal relative to the instrument noise. We
describe a simple method for removing the dominant, Doppler, component of the signal modulation. The
demodulation reassembles the power from a monochromatic source into a narrow spike, and provides
a quick way to determine the sky locations and frequencies of the brightest gravitational
wave sources.
\end{abstract}

\maketitle

\section{Introduction}

This paper introduces a ``quick and dirty'' method for making a first pass at the LISA data analysis problem.
LISA will be sensitive to gravitational wave sources in the frequency range $10^{-4}$ to $10^{-1}$ Hz, which
nicely complements the frequencies range covered by the ground based detectors ($10^1$ to $10^3$ Hz).
The data analysis challenges posed by LISA are very different from those encountered with the
ground based detectors. Unlike the situation faced by the ground based observatories,
most of the sources that LISA hopes to detect have well understood gravitational waveforms - indeed
many of them will be monochromatic and unchanging over the life of the experiment. The difficulty comes
when we include LISA's orbital motion and the modulations this introduces into the signal. Since LISA
is designed to orbit at 1 AU, the modulations introduce sidebands at multiples of the orbital frequency
of $f_m = 1/{\rm year}$.


The approach we investigate here works by correcting for the dominant, Doppler modulation. This has the
effect of re-assembling most of the Fourier power of a monochromatic source into a single spike. Since
the correction depends on where the source is located on the sky, we are able
to locate a source and its frequency.

The Doppler demodulation approach is familiar to radio astronomers, and has been discussed in relation to
gravitational wave astronomy\cite{livas}. Doppler demodulation forms an integral part of ground based
gravitational wave searches for continuous gravitational wave sources such as pulsars\cite{pb,dv}. However,
the unique orbital motion of the LISA observatory, and the different frequency range that it covers,
demand a separate treatment of the Doppler demodulation approach to LISA data analysis.

\section{Signal Modulation}

LISA will output a time series describing the strain in the detector as a function of time, $h(t)$.
The strain will be a combination of the signal, $s(t)$, and noise, $n(t)$, in the detector: $h(t) = s(t)
+ n(t)$. Because of LISA's orbital motion, a source that is monochromatic at the Sun's barycenter
will be spread over a range of frequencies. A monochromatic gravitational wave with frequency $f$
and amplitudes $h_+$ and $h_\times$ in the plus and cross polarizations will produce the response\cite{curt}
\begin{equation}
s(t) = A(t) \cos \Psi(t)
\end{equation}
where
\begin{equation}
\Psi(t)=\left[ 2\pi f t +\varphi_0 + \phi_D(t) + \phi_P(t) \right]\, .
\end{equation}
The amplitude modulation $A(t)$, frequency modulation $\phi_D(t)$ and
phase modulation $\phi_P(t)$ are given by
\begin{eqnarray}
A(t) &=& \left[ (h_+ F^+(t))^2+(h_\times F^\times (t))^2 \right]^{1/2} \\
\nonumber \\
\phi_D(t) & = & 2\pi f \frac{R}{c}\sin\bar{\theta}_s \cos(2\pi f_m t - \bar \phi_s) \\
\nonumber \\
\phi_P(t) & = & - {\rm arctan}\left(\frac{h_\times F^\times(t)}{h_+ F^+(t)}\right) \, .
\end{eqnarray}
Here $F^+(t)$ and $F^\times (t)$ are the detector antenna patterns in barycentric coordinates\cite{curt}.
Each of the modulations is periodic in the orbital frequency $f_m=1/{\rm year}$. The angles $\bar{\theta}_s$
and $\bar \phi_s$ give the sky location of the source, $\varphi_0$ is the phase of the wave and $R=1$ AU is the
Earth-Sun distance. The frequency modulation, which is caused by the time dependent Doppler shifting of the
gravitational wave in the rest frame of the detector, is the dominant effect for frequencies above
$5 \times 10^{-4}$ Hz\cite{ns}. Ignoring the amplitude and phase modulation for now, 
pure Doppler modulation takes the form
\begin{eqnarray}\label{dop}
&& s(t) = A \cos\left[2 \pi f t + \beta \cos(2\pi f_m t-\bar \phi_s) +\varphi_0 \right] \nonumber \\
  && \quad = \Re\left( A  \sum_{n=-\infty}^{\infty} J_n(\beta) e^{2\pi i (f+f_m n) t}
e^{i \varphi_0}e^{i n (\pi/2-\bar \phi_s)}\right).
\end{eqnarray}
Here $J_n$ is a Bessel function of the first kind of order $n$ and $\beta$ is the modulation index
\begin{equation}
\beta = 2\pi f \frac{R}{c} \sin\bar\theta_s \, .
\end{equation}
The bandwidth of the signal - defined to be the frequency interval that contains 98\% of the total power -
is given by
\begin{equation}
B = 2(1+\beta) f_m \, .
\end{equation}
Sources in the equatorial plane have bandwidths ranging from $B = 2.6 \times 10^{-7}$ Hz at
$f=10^{-3}$ Hz to $B = 2.1 \times 10^{-6}$ Hz at $f=10^{-2}$ Hz.

\section{Demodulation}

If LISA were at rest with respect to the sky, each monochromatic source would produce
a spike in the power spectrum of $s(t)$. While this would make data analysis very easy, it
would also severely limit the science that could be done as it would be impossible to
determine the source location or orientation. Since LISA will move with respect to the
sky, each source will have its own unique modulation pattern, and this pattern can be
used to fix its location and orientation. On the other hand, the modulation makes the
data analysis more complicated as the Fourier power of each source is spread over a
wide bandwidth $B$. One approach to the data analysis problem is to
demodulate the signal, thereby re-assembling all the power of a given source at one
frequency. Since each source has a unique modulation, each source will also require
a unique demodulation. Demodulating a particular source is not difficult if one happens
to know its frequency, location, orientation and orbital phase. One could imagine searching
through this parameter space and looking for spikes in the power spectrum corresponding
to sources that have been properly demodulated. A more practical approach is to focus
on the Doppler modulation as it causes the largest spreading of the Fourier power.

Consider the Doppler modulated phase
\begin{equation}
\Phi(t) = 2\pi f \left[t+ \frac{R}{c}\sin\bar{\theta}_s
\cos(2\pi f_m t - \bar \phi_s)\right] +\varphi_0 \, .
\end{equation}
We seek a new time coordinate $t'$ in which this phase is stationary:
\begin{equation}
\frac{d \Phi}{d t'} = 2\pi f = \frac{d \Phi}{d t} \frac{ d t}{d t'} \, .
\end{equation}
Thus,
\begin{equation}
t' = \int^t \left(1 - \frac{v}{c} \sin\bar \theta_s \sin(2\pi f_m t - \bar\phi_s)\right) dt \, .
\end{equation}
Working to first order in $v/c$ we have
\begin{equation}\label{ctran}
t = t' - \frac{R}{c} \sin\bar{\theta}_s \cos(2\pi f_m t - \bar \phi_s)
\end{equation}

Taking the data stream from the detector over a one year period and
performing a fast Fourier transform allows us to write
\begin{equation}
s(t) = \sum_n a_n e^{2 \pi i f_m n t} \, .
\end{equation}
Performing the coordinate transformation (\ref{ctran}) we arrive at the new
Fourier expansion
\begin{equation}
s(t') = \sum_k c_k e^{2 \pi i f_m k t'} \, ,
\end{equation}
where the Fourier coefficients $c_k$ are given by
\begin{equation}
c_k = \sum_n a_n J_{n-k}(2\pi n f_m \frac{R}{c} \sin\bar\theta_s) e^{i(n-k)(\bar\phi_s-\pi/2)} \, .
\end{equation}
Since the modulation has a limited bandwidth,
an excellent approximation to $c_k$ is given by
\begin{equation}
c_k \simeq \sum_{n=k-l}^{k+l} a_n J_{n-k}(\alpha) e^{i(n-k)(\bar\phi_s-\pi/2)} \, ,
\end{equation}
where
\begin{equation}
\alpha = 2\pi k f_m \frac{R}{c} \sin\bar\theta_s
\end{equation}
and
\begin{equation}
l = 1 + [\alpha] \, .
\end{equation}
The square brackets imply that we have taken the integer part of $\alpha$. Consider the
pure Doppler modulation described in (\ref{dop}) for
the simple case where $f=q f_m$, and $q$ is an integer (the non-integer case is more involved, but
the idea is the same). The source has Fourier components
\begin{equation}
a_n = A \, e^{i \varphi_0} J_{n-q}(\beta) e^{i(n-q)(\pi/2-\bar\phi_s)}\, ,
\end{equation}
so that
\begin{equation}
c_k \simeq A\,   e^{i \varphi_0} e^{i(q-k)(\bar\phi_s-\pi/2)} \sum_{n} J_{n-q}(\beta) J_{n-k}(\alpha)  \, .
\end{equation}
Using the Neumann addition theorem,
\begin{equation}
\sum_{n} J_{n-q}(\beta) J_{n-k}(\alpha) =  J_{k-q}(\alpha-\beta),
\end{equation}
and the fact that $\alpha \approx \beta$, we find
\begin{equation}
c_k \simeq A \,  e^{i \varphi_0} \delta_{kq} \, .
\end{equation}
In other words, the Doppler demodulation procedure re-assembles all the power at the barycenter frequency
$f=q f_m$. The demodulation is less effective when applied to a real LISA source because it does not
correct for the amplitude and phase modulations.

Our Fourier space approach is very efficient if one is interested in a limited frequency range.
It does not offer any great saving over a direct implementation in the time domain if one wants
to consider all frequencies at once. Hellings\cite{ron} has
recently implemented the Doppler demodulation procedure in the time domain and
found similar results to ours.

\begin{figure}[t]
\vspace{60mm}
\includegraphics{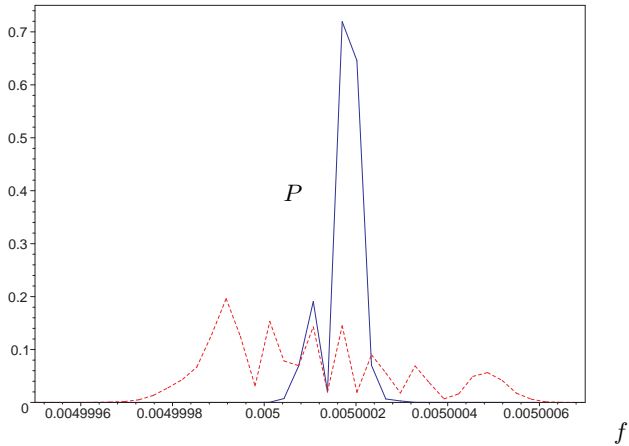}
\vspace{5mm}
\caption{Power spectra, $P(f)$, before (dashed line) and after (solid line) Doppler demodulation.}
\begin{picture}(0,0)
\put(0,135){$P$}
\put(125,45){$f$}
\end{picture}
\end{figure}

To illustrate how the Doppler demodulation works we consider monochromatic, circular Newtonian binaries
as described in Ref.~\cite{curt}. Figure 1 shows the power spectrum of $s(t)$ before and
after Doppler demodulation for a source with $f= 5.000167 \times 10^{-3}$ Hz,
$\bar \theta_s = 2.385366$ and $\bar \phi_s = 4.462868$. We see that most of the power is
collected into a spike of width $\sim 3 f_m$ about the barycenter frequency.

\begin{table}
\begin{tabular}{|c|c|c|c|}
\hline
\hspace*{0.2in} & \hspace*{0.25in} $f$ (Hz) \hspace*{0.25in} & \hspace*{0.25in} $\bar \theta_s$ \hspace*{0.25in} 
& \hspace*{0.25in} $\bar \phi_s$  \hspace*{0.25in} \\
\hline
A & $4.999904\times 10^{-3}$  & 2.867424 & 0.258971 \\
B & $4.999729\times 10^{-3}$  & 0.661738  &  3.323842 \\
\hline
\end{tabular}
\caption{Description of Sources A and B}
\end{table}

Next we consider an example where there are two sources that are nearby in frequency, but
nevertheless do not have overlapping bandwidths. The sources have $f=5.001419\times 10^{-3}$ Hz,
$(\bar \theta_s, \bar \phi_s) = (2.056685, 2.212044)$ and $f=4.9988\times 10^{-3}$ Hz,
$(\bar \theta_s, \bar \phi_s) = (1.280671, 1.446971)$. The higher frequency source has an amplitude
$1.103$ times larger than the lower frequency source. The effect of demodulating each source
is shown in Figure 2.

\begin{figure}[ht]
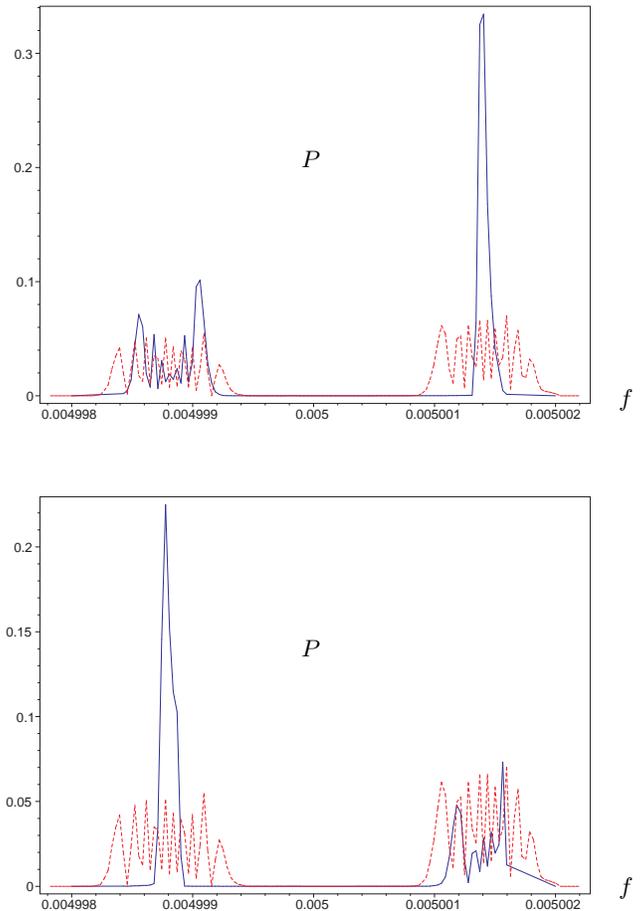

\vspace{120mm}
\includegraphics{demod1.ps}
\includegraphics{demod2.ps}
\vspace{5mm}
\caption{Power spectra before (dashed line) and after (solid line) Doppler demodulation.
The upper graph shows the demodulation when the detector is made stationary with respect to the higher frequency
source, while the lower graph shows the same for the lower frequency source.}
\begin{picture}(0,0)
\put(5,360){$P$}
\put(125,270){$f$}
\put(5,175){$P$}
\put(125,85){$f$}
\end{picture}
\end{figure}

The performance of the demodulation procedure is considerably less impressive when the two sources
have overlapping bandwidths. Consider sources A and B described in Table 1.
The amplitude of Source B is 1.59 times larger than
the amplitude of Source A. Figure 3 shows what happens when we demodulate Source A and Source B in turn.
Because the two sources overlap, their signals interfere with each other, and the demodulation is only able
to detect the stronger of the two sources.

\begin{figure}[ht]
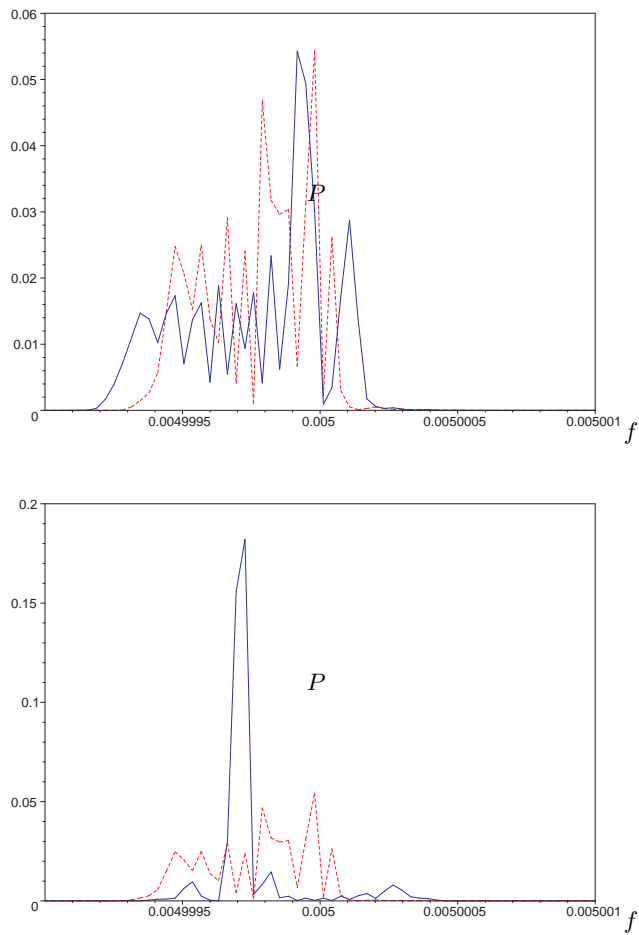

\vspace{120mm}
\includegraphics{idemod1.ps}
\includegraphics{idemod2.ps}
\vspace{5mm}
\caption{Power spectra before (dashed line) and after (solid line) Doppler demodulation.
The upper graph shows the demodulation when the detector is made stationary with respect to Source A,
while the lower graph shows the same for Source B.}
\begin{picture}(0,0)
\put(5,350){$P$}
\put(125,260){$f$}
\put(5,165){$P$}
\put(125,75){$f$}
\end{picture}
\end{figure}

\subsection{Angular resolution}

\begin{figure}[ht]
\vspace{43mm}
\includegraphics{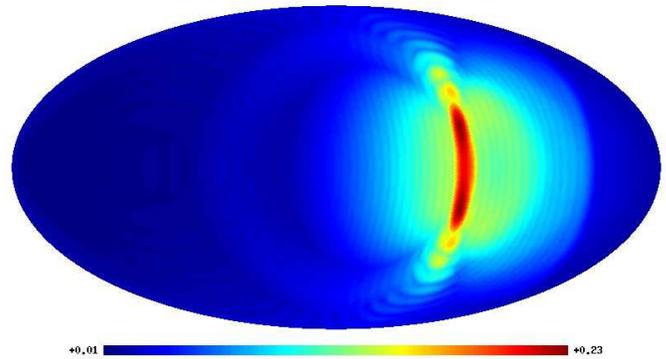}
\vspace{8mm}
\caption{Determining the sky location of a source with $(\bar \theta_s, \bar \phi_s) = (1.851478, 4.934289)$.
The sky map uses the Mollweide projection in Ecliptic coordinates with $(\pi/2,0)$ at the center of the
map.}
\end{figure}

The Doppler demodulation technique can be used to find the sky location of a source by successively
demodulating each point on the sky. In practice we use the HEALPIX\cite{kris} hierarchical,
equal area pixelization scheme to provide a finite number of sky locations to demodulate. We
search for the maximum power contained in three adjacent frequency bins, and record this value
for each sky pixel. Rather than find the maxima across all frequencies, we produce separate sky
maps for frequency intervals of width $4\pi f \frac{R}{c} f_m$. There is little point in
trying to use smaller frequency intervals due to the interference effect discussed earlier.

\begin{figure}[ht]
\vspace{43mm}
\includegraphics{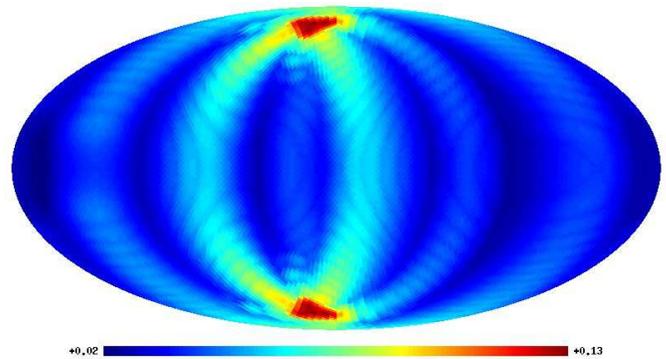}
\vspace{8mm}
\caption{Determining the sky location of source A without source B.}
\end{figure} 

We begin by considering an isolated source ({\rm i.e.} no other sources with overlapping bandwidths) with
$f=5.00015\times 10^{-3}$ Hz and $(\bar \theta_s, \bar \phi_s) = (1.851478, 4.934289)$. Using sky pixels
of angular size $\sim 0.92^o$ we arrive at the graph shown in Figure 4. The demodulation code produced the
best fit values of $f=5.00014 \times 10^{-3}$ Hz and $(\bar \theta_s, \bar \phi_s) = (1.253, 5.04)$ and
$(\bar \theta_s, \bar \phi_s) = (1.889, 5.04)$. The degenerate fit for the sky location illustrates one of
the drawbacks of the Doppler demodulation method - it is unable to distinguish between sources above or
below the equator. The errors in the source location, $(\Delta \bar \theta_s = 2.1^o, \Delta \bar \phi_s = 6.3^o)$,
are larger than the pixel size, and can be attributed to the amplitude and phase modulations which we
have not corrected for. It should be noted that the error in the source location is not caused by
instrument noise since we are working in the large signal-to-noise limit ($n(t)=0$). Indeed, instrument noise
has little effect on the demodulation procedure since the noise is incoherent. The noise gets moved about
in frequency space, but there is little power accumulation at any one frequency. 
When noise was added to the previous example at a signal to noise ratio
of 5, the frequency determination and $\bar \phi_s$ determination were unaffected, while the error in
$\bar \theta_s$ grew to $\Delta \bar \theta_s = 3.4^o$.

\begin{figure}[ht]
\vspace{43mm}
\includegraphics{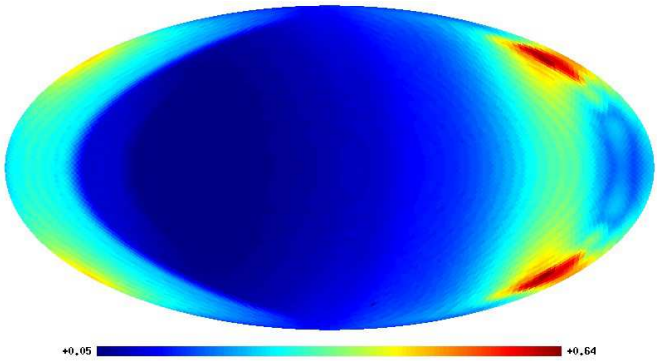}
\vspace{5mm}
\caption{Determining the sky location of source B without source A.}
\end{figure}

Finally, we illustrate how the source interference phenomena discussed earlier limits our ability to
locate sources that overlap in frequency. Figures 5 and 6 show how the demodulation procedure is able
to locate Sources A and B when one of the sources is turned off. Figure 7 shows what happens when
both sources are present. We see that source B shows up clearly, while source A is washed out.

\begin{figure}[ht]
\vspace{43mm}
\includegraphics{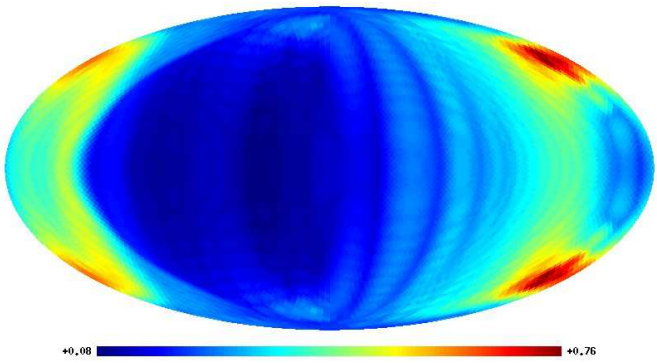}
\vspace{5mm}
\caption{Determining the sky location of sources A and B together.}
\end{figure}

\section{Discussion}

The Doppler demodulation procedure provides a quick way of finding the frequencies and sky locations
of the brightest sources detected by LISA. The method has a number of limitations, the most serious being
its inability to locate more than one source per bandwidth, and its inability to determine if a source
is in the northern or southern hemisphere. Despite these limitations, the Doppler demodulation procedure
will be a useful tool in the LISA data analysis arsenal.

\section*{Acknowledgements}

It is a pleasure to thank Tom Prince and the members of the Montana Gravitational Wave Astronomy Group
- Bill Hisock, Ron Hellings and Matt Benacquista - for many stimulating discussions. The work of N.\ J.\ C.\
was supported by the NASA EPSCoR program through Cooperative Agreement NCC5-579. The work of S.\ L.\ L. was 
supported by LISA contract number PO 1217163.

\end{document}